\RequirePackage{rotating}
\documentclass[useAMS]{mn2e}
\usepackage{mhchem}
\usepackage{graphicx}
\usepackage{longtable}
\usepackage{rotate}
\usepackage{rotating}
\usepackage{mathtools}
\usepackage{pdflscape}
\usepackage{scalerel}
\usepackage{times}
\usepackage{verbatim}
\newif\ifAMStwofonts

\def\gs{\mathrel{\hbox{\rlap{\hbox{\lower4pt\hbox{$\sim$}}}\hbox{$>$}}}}
\def\ls{\mathrel{\hbox{\rlap{\hbox{\lower4pt\hbox{$\sim$}}}\hbox{$<$}}}}

\def\hitomi{{\it Hitomi}}
\def\athena{{\it Athena}}
\def\xrism{{\it XRISM}}

\def\asca{{\it ASCA}}

\def\xmm{{\it XMM-Newton}}

\def\et{{et al.\ }}
\def\mcg{{MCG--6-30-15}}
\def\mrk335{{Mrk~335}}
\def\rg{{\thinspace r_{\rm g}}}

\def\feka{{Fe~K$\alpha$}}

\def\fekb{{Fe~K$\beta$}}

\def\deg{^{\circ}}

\def\cm{{\rm\thinspace cm}}
\def\erg{{\rm\thinspace erg}}

\def\keV{{\rm\thinspace keV}}

\def\s{{\rm\thinspace s}}
\def\ks{{\rm\thinspace ks}}

\def\cmps{\hbox{$\cm\s^{-1}\,$}}

\def\ergpscmps{\hbox{$\erg\cm^{-2}\s^{-1}\,$}}

\title[Nuclear spallation in active galaxies]
      {
Nuclear spallation in active galaxies    }

\author[L. C. Gallo \et]
       {
       L. C. Gallo,$^1$ 
       J. S. Randhawa,$^{1,2}$
       S. G. H. Waddell,$^1$
       M. H. Hani,$^{1,3}$\thanks{Vanier Scholar}
       J. A. Garc\'ia,$^{4,5}$ 
\newauthor      and  C. S. Reynolds$^6$
        \\ 
$^{1}$ Department of Astronomy and Physics, Saint Mary's University, 923 Robie Street, Halifax, NS, B3H 3C3, Canada \\
$^{2}$ National Superconducting Cyclotron Laboratory, Michigan State University, East Lansing, MI 48824, USA \\
$^{3}$ Department of Physics and Astronomy, University of Victoria, Victoria, British Columbia, V8P 1A1, Canada  \\
$^{4}$ Cahill Center for Astronomy and Astrophysics, California Institute of Technology, Pasadena, CA 91125, USA \\
$^{5}$ Dr. Karl Remeis-Observatory and Erlangen Centre for Astroparticle Physics, Sternwartstr.~7, 96049 Bamberg, Germany \\
$^{6}$ Institute of Astronomy, University of Cambridge, Madingley Road, Cambridge, CB3 0HA, United Kingdom \\
}
\date{Accepted. Received. }
\pagerange{\pageref{firstpage}--\pageref{lastpage}}
\pubyear{2010}
\begin{document}
\maketitle
\label{firstpage}

\begin{abstract}
A number of works point to the presence of narrow emission features at unusual energies in the X-ray spectra of active galactic nuclei (AGN) or to potentially low iron abundances as possible evidence for the spallation of iron.  With the imminence of high-resolution calorimeter spectroscopy, the potential to test spallation models in astrophysical sources will soon be possible.  Previously determined nuclear spallation reactions of Fe are recalculated making use of improved total inelastic and partial reaction cross sections that result in different absolute and relative abundances of the main spallation elements Mn, Cr, V, and Ti.  The effects of ionisation and dynamics near the black hole are examined in simulated spectra with CCD and calorimeter (i.e. \hitomi) resolution.  With high-resolution, differences in relative abundances and ionisation should be detectable if spallation is originating at large distances from the black hole (e.g. torus or disc wind) where blurring is not significant.  If spallation were occurring in the inner accretion disc, it would likely be undetected as blurring effects would cause significant blending of spectral features.
\end{abstract}



\section{Introduction}
\label{sect:intro}

The bombardment of cosmic rays (highly energetic protons and atomic nuclei) on matter can lead to the formation of elements through spallation.  As the high-energy, charged particles strike atomic nuclei, nucleons can be stripped from the atom, thereby forming isotopes and lighter elements, and altering the overall abundances.  The relatively high abundance of iron in cosmic matter renders the metal a natural target for spallation.  Skibo (1997, hereafter S97) found that the spallation of Fe will result in its depletion, and the enhancement of Ti~K$\alpha$, V~K$\alpha$, Cr~K$\alpha$, and Mn~K$\alpha$ between $\sim4.5-6\keV$ in the X-ray band.

The importance of spallation in active galactic nuclei (AGN) is often debated and there is yet clear evidence for it.  A number of works point to the presence of narrow emission features at unusual energies in the X-ray spectra of AGN (e.g. Xu \et 2016; Lobban \et 2011; Turner \et 2010, 2002; Turner \& Miller 2010; Della Ceca \et 2005; Porquet \et 2004; Gallo \et 2004) or to potentially low iron abundance (e.g. Bonson \et 2015) as possible evidence for spallation.  Often, the  interpretation is dismissed because the line energies and/or strengths are inconsistent with predictions.   S97 proposed that the broad emission profile seen in some AGN with \asca\ could be attributed to spallation viewed with the moderate CCD spectral resolution of the instrument rather than to a relativistically broadened \feka\ emission line.  However, spallation cannot mimic relativistic broad lines.  Even with moderate resolution CCD detectors, spallation should be distinguishable from broad lines (see Sect.~\ref{sect:spec}).

During its brief operation, the calorimeter mission  \hitomi\ (Takahashi \et 2016) demonstrated the massive progress that will be achieved with high-resolution X-ray spectroscopy (e.g. Hitomi Collaboration 2016, 2017, 2018).  With future calorimeter missions like \athena\ (Barret \et 2018) and \xrism\ (Tashiro \et 2018) in development, the potential to examine spallation in astronomical sources will become reality.  In this work, we examine the appearance of spallation in the reflection spectrum of AGN.  The effects of ionisation on the spalled elements are taken into account, as are the dynamics of the emitting material in the vicinity of the black hole.  The potential to detect spallation signatures with current CCDs and with future calorimeters is also presented.  In the next section, the spallation abundances originally calculated by S97 are refined by making use of improved cross sections.


\section{Spallation and abundance evolution}
\label{sect:spallation}

Nuclear spallation is a process in which a light projectile (proton, neutron or light particle) with kinetic energy of several MeV to GeV interacts with a nucleus leading to the emission of a large number of hadrons (mainly neutrons) or fragments. S97 assumed that in galactic centres, the central source produces the energetic protons with luminosity equivalent to that of emerging radiation based on equipartition of energy near the central engine. The bombardment of accreted material with these high energy protons lead to the spallation of heavier elements and hence alters the composition of accreted material. 

Here the methodology of S97 is followed to obtain the abundances of Fe and sub-Fe elements (mainly Mn, Cr, V and Ti), using improved total inelastic and partial reaction cross sections wherever possible.  As in S97, the mean rate per target nucleus for nuclear spallation from species $i$ to $j$ is given by:
\begin{equation}  
R_{ij}(E)=4\pi\int_{0}^{\infty }\sigma _{ij}(E)J_{p}(E)\mathrm{d}E
\end{equation}
where $\sigma_{ij}$ is the appropriate spallation cross section and $J_{p}$ is the steady state proton intensity (protons s$^{-1}$ cm$^{-2}$ sr$^{-1}$ MeV$^{-1}$).

\begin{figure}
\includegraphics[width=\columnwidth]{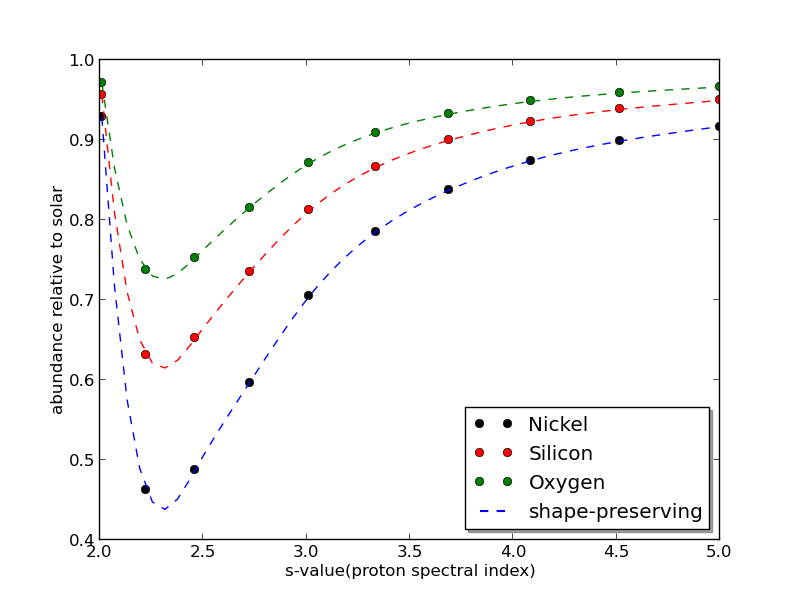}
\caption{Abundance evolution for Ni, Si and O as a function of injected energy spectral index.}
\label{fig:abund}
\end{figure}

S97 assumed the injection spectrum followed a simple
power law in kinetic energy so that $J_{p}$ was proportional to the integral:
\begin{equation}
\frac{\mathrm{d} N_{p}(E)}{\mathrm{d} t \mathrm{d}E} =(s-2)\eta\dot{M}c^{2}E_{o}^{s-2}E^{-s}
\end{equation}
where the spectral index was $s>$2. In this work, we find the abundance changes are maximum at $s=2.3$ (e.g. see Fig~\ref{fig:abund}), therefore, all the abundances have been reported at $s=2.3$, unless stated otherwise. The accretion efficiency is $\eta =0.1$, $E_{0}$ is the low-energy cut-off (10 MeV), and $E$ is the proton energy in MeV.

The system of equations regulating the abundances of Fe and the production of sub-elements is: 
\begin{equation}
N_{Fe}(t)= N_{Fe}^{\bigodot}e^{-R_{Fe} t}
\end{equation}
\begin{multline}
N_{i}(t)= N_{i}^{\bigodot}e^{-R_{i} t} 
+N_{Fe}^{\bigodot}(e^{-R_{Fe} t}+e^{-R_{i} t}) \\
\times (R_{Fe\rightarrow i}+\sum_{j}R_{Fe\rightarrow j(i)})/(R_{i}+R_{Fe})    
\end{multline}

In Equation 3 and 4, $N^{\bigodot}$ represents the solar abundance of the element and $t$ is the accretion time scale. 
Equation 3 dictates the change in the iron abundance due to the destruction into all the possible sub-iron elements. This is referred to as the total destruction mode and the rate ($R_{Fe}$) is dependent on the total inelastic cross sections (see Equation 1).   As in S97, the total inelastic cross sections used are from Letaw \et (1983).  However, in this work an additional scaling factor is used at $\sim20$~MeV to better match the experimental data of Tripathi \et (1997).

Equation 4 provides the abundance evolution of the $i^{th}$ element, where the first term represents its destruction, and the other terms govern its production due to the spallation of all  heavier elements, up to Fe.  Here, the rates $R_{Fe\rightarrow i}$ and $R_{Fe\rightarrow j(i)}$ represent the
partial rate of Fe being destroyed to some element $i$ directly or through the formation of an unstable isotope $j$ that then decays to $i$, respectively. 

These partial rates use partial cross sections based on the semi-empirical formulae from Silberberg \et (1981) and those estimates have  been updated many times using scaling factors based on different experimental results (Silberberg \et 1977, 1990, 1998).   In this work, we  have used the routines written by A.F. Barghouty, which include updates of the cross sections in Silberberg \et (1998) and that are adopted by the GALPROP\footnote{https://galprop.stanford.edu/index.php} research group to calculate propagation of cosmic rays in galaxies (Moskalenko \et 2003). These routines implement a major improvement in the cross sections of the atomic mass elements of interest for this work.   For example, the  production of V and Ti from the targets Cr- Ni (i.e. Z$_{t}$= 24-28) changes by a factor of two.   
\begin{figure}
\includegraphics[width=\columnwidth]{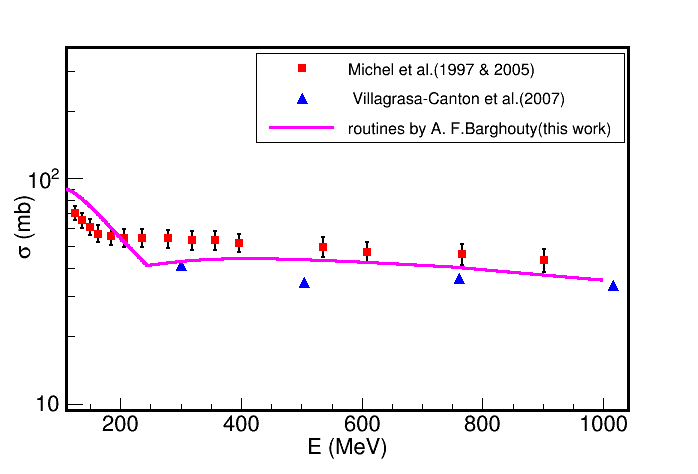}
\caption{The experimental cross section for \ce{p + Fe ->51Cr} obtained from direct kinematic measurements (Michel \et 1997) and from inverse kinematic measurements (Villagrasa-Canton \et 2007). The theoretical cross sections from the routines adopted in this current work are shown as the solid curve. }
\label{fig:cross}
\end{figure}
\begin{figure}
\includegraphics[width=\columnwidth]{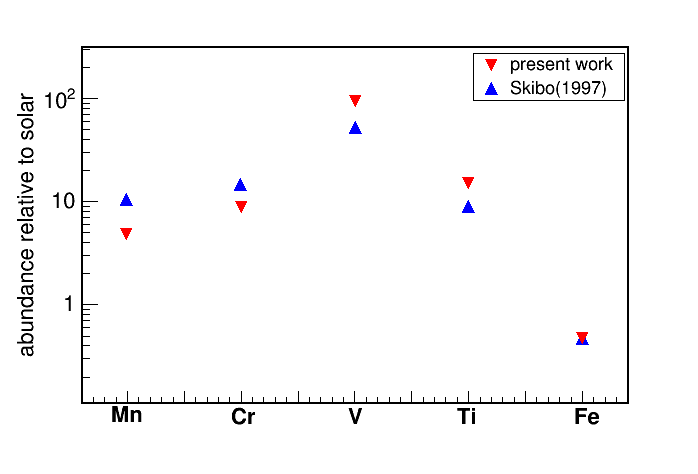}
\caption{Comparison of Fe, Mn, Cr, V, Ti abundances obtained in this work to those from Skibo (1997). As the methodology is adopted from S97, the difference in abundances originates from the difference in cross sections. The total inelastic cross section, relevant to the destruction of Fe, is similar to that used in S97, hence the Fe abundance is the same as S97.}
\label{fig:comp}
\end{figure}
\begin{figure}
\includegraphics[width=\columnwidth]{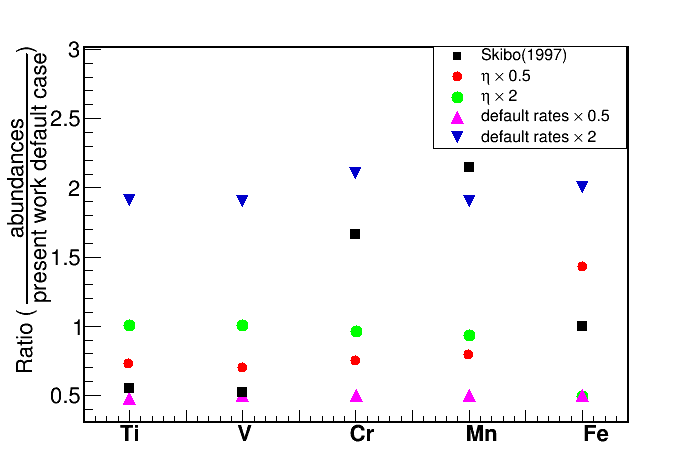}
\caption{The effects on the abundances from changing the accretion efficiency ($\eta$, filled circles) and cross sections (filled triangles) by an arbitrary factor of two.  Uncertainties in cross sections can be significant on the resulting spallation abundances, but changes in $\eta$ are less important.}
\label{fig:errors}
\end{figure}

The main advantage of  using these routines over the other existing parameterizations  is that in addition to the energy dependent scaling, the predicted cross sections agree with the experimental data  from the work of Michel \et (1997), which is one of the primary works in the measurement of proton induced spallation of Fe. Note that an additional scaling factor was used in the calculations wherever required to match with the experimental data. Displayed in Fig.~\ref{fig:cross} is the experimental cross section for \ce{p + Fe ->51Cr} obtained from direct kinematic measurements (Michel \et 1997) and from inverse kinematic measurements (Villagrasa-Canton \et 2007). The theoretical cross sections from the routines adopted in this  current work are shown in comparison. 

Using these updated cross sections and the methodology of S97, the abundance of Mn, Cr, V and Ti obtained in this work are compared to those of S97 in Fig.~\ref{fig:comp}. Compared to S97, the abundance (relative to solar) for Mn and Cr is lower, wheres V and Ti are produced in greater quantity. 

The other possible observational signature of spallation in AGN X-ray spectra could come from the destruction of the elements other than iron. There is no reason why spallation should be limited to iron as any element with relatively high abundance in the accretion disc will be a natural target for the proton induced spallation. As pointed earlier, the total spallation induced destruction depends on the total inelastic cross sections which varies with the atomic mass number ($A$) as $A^{2/3}$. This means that if spallation is present in AGN, the ratio of prominent element lines will scale as $A^{2/3}$ when normalized to abundance. As an example, the abundance of three other relatively abundant elements i.e. Ni, Si and O is shown as a function of spectral index (Fig.~$\ref{fig:abund}$). At $s= 2.3$, the change in abundance is maximum and line ratios could infer if spallation is indeed present.  

Even though the default cross sections in this work are made to agree with the available experimental data, the lack of experimental data for many isotopes introduces some uncertainty in the cross sections. To begin assessing the impact on the absolute abundances, a factor of two uncertainty in all the cross sections is assumed and the effect is shown in Fig.~\ref{fig:errors}.  Other uncertainties in the current calculations will arise from the unknown intensity of incident protons, which depends on the spectral index ($s$) and the accretion efficiency ($\eta$). The proton spectrum is chosen to match the source spectrum of Galactic cosmic rays (Berezinskii et al. 1990) and the spectral index is chosen to be $s=2.3$ to maximize the effects of spallation. The accretion efficiency was varied by a factor of two to determine the impact on measured abundances. The results are shown in Fig.~\ref{fig:errors}. The abundances are not very sensitive to changes in $\eta$, but uncertainties in cross sections can result in significant differences in abundances.

\section{The X-ray spectrum}
\label{sect:spec}
\begin{figure*}
\begin{center}
\begin{minipage}{0.48\linewidth}
\scalebox{0.32}{\includegraphics[angle=0,trim= 1.8cm 1cm 2.7cm 2cm, clip=true]{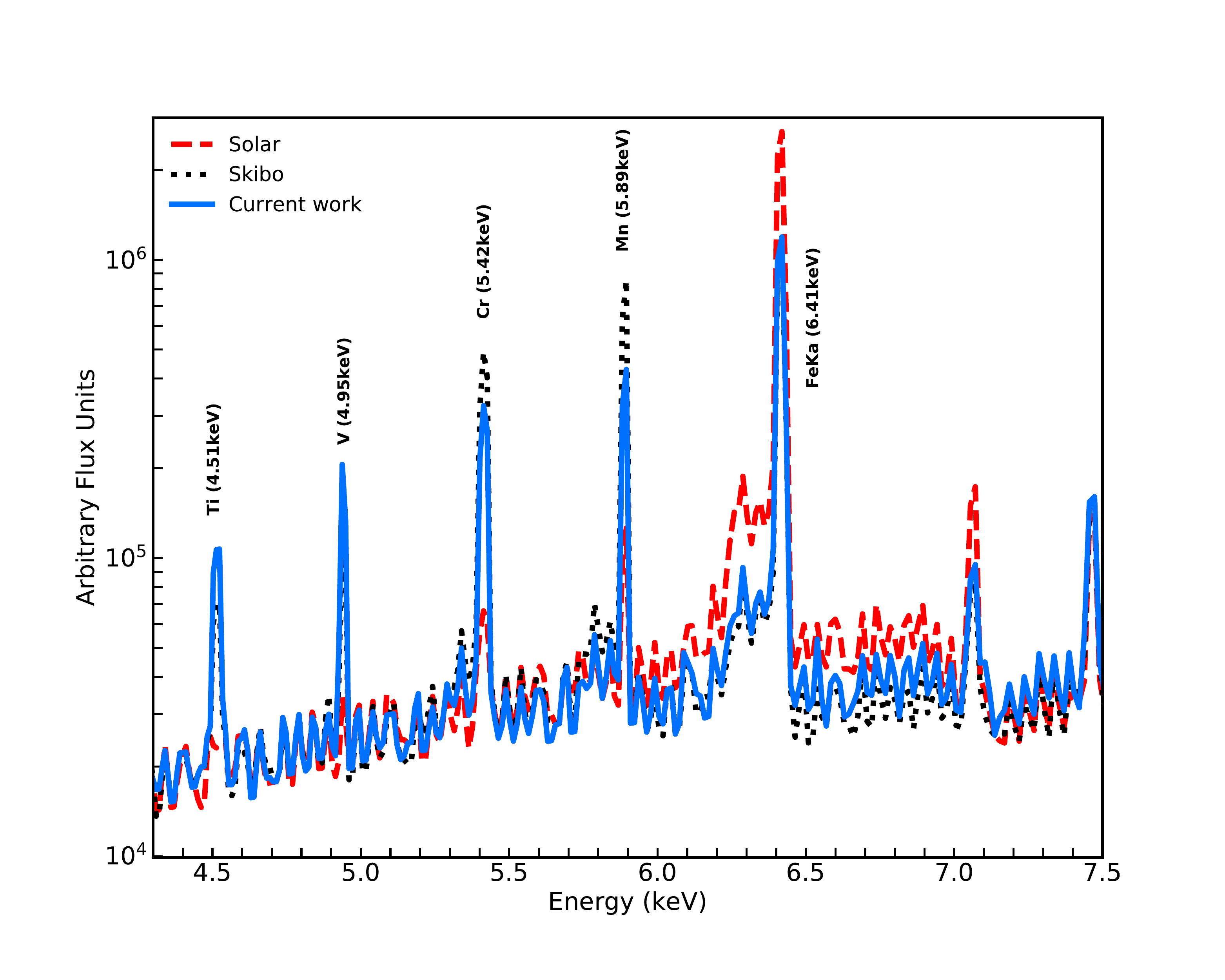}}
\end{minipage}  \hfill 
\begin{minipage}{0.48\linewidth}
\scalebox{0.32}{\includegraphics[angle=0,trim= 1.8cm 1cm 2.7cm 2cm, clip=true]{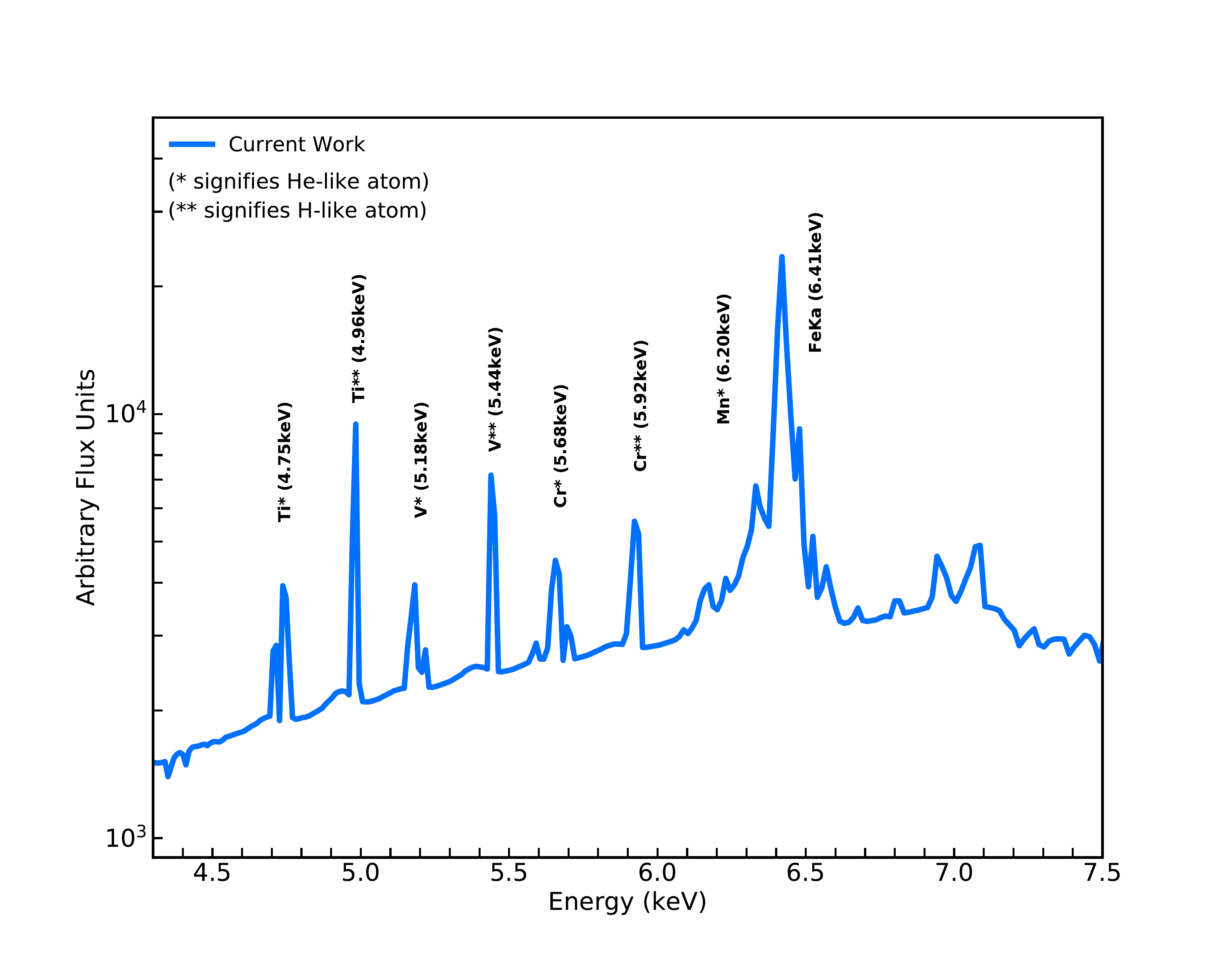}}
\end{minipage}
\end{center}
\caption{Left: The reflection spectrum from neutral material with solar abundances (red, dashed curve) compared to spectra with modified abundances predicted by different spallation models.  The abundances from Skibo (1997, S97) are shown as the black, dotted curve while those from this work (Sect.~\ref{sect:spallation}) are shown as the blue, solid curve. 
Right: The reflection spectrum from ionised material ($\xi = 100\erg\cmps$) using the spallation abundances calculated in this work (Sect.~\ref{sect:spallation}).  Note, that the adopted ionisaton model ({\sc xillver}), currently does not include the neutral stages of these Fe-peak elements, therefore only He- and H-like transitions are evident in the figure.}
\label{fig:jaspreet}
\end{figure*}

The appearance of cosmic spallation from optically thick material that is illuminated by a power law X-ray source is considered.  The so-called reflection spectrum is associated with the accretion disc and the torus in AGN.  The primary difference is that the distant torus is likely cold (neutral) and motions are small, whereas the material in the accretion disc can be ionised and subject to high rotational velocities and gravitational effects.

To investigate the presences of cosmic spallation in neutral reflection, the Monte Carlo simulation of X-ray reflection by Reynolds \et (1995, 1994; see also George \& Fabian 1991) is adopted.  The original model already includes K$\alpha$ fluorescence of C, O, Ne, Mg, Si, S, Ar, Ca, Cr, Fe and Ni, but the code had to be modified to include the elements Mn, V, and Ti.  The three elements are added to the model assuming solar abundance from Anders \& Grevesse (1989) and the fluorescent yields from Bambynek \et (1972).  To visualize the effects of spallation, the abundances of Mn, V, Ti, Cr, and Fe are altered accordingly (see S97 and Sect.~\ref{sect:spallation}, Fig.~\ref{fig:comp})  The medium is illuminated by an X-ray source with a photon index of $\Gamma=1.9$ and viewed at an inclination of $45\deg$. 

In Fig.~\ref{fig:jaspreet} (left panel), the appearance of cold reflection is compared for material with solar abundances (red dashed curve), and for abundances modified by cosmic spallation.  As noted above, the spallation model calculated in this work (Sect.~\ref{sect:spallation}) generates weaker Cr and Mn emission, but stronger Ti and V emission than was predicted by S97.  While the absolute line fluxes show little differences from S97, the line ratios like $\frac{\rm Cr}{\rm V}$ are substantially different between models. The weaker \feka\ and \fekb\ are comparable in both models since the total inelastic cross section, relevant to the destruction of Fe, is similar to that used in S97.

In ionised material, the significance of spallation could be more important as more atomic transitions become possible.  The X-ray reflection code {\sc xillver} (Garc\'ia \et 2010, 2013) is used to examine the appearance of the emission lines from ionised material undergoing spallation.  As in the neutral case, the simulated plasma is illuminated by a power law with a photon index $\Gamma=1.9$ and cutoff energy of $E_{cut} = 300\keV$.  The ionisation parameter ($\xi=4\pi F/n$ where $n$ is the hydrogen number density and $F$ is the incident flux) is held at $\xi = 100\erg\cmps$.    Although {\sc xillver} uses abundances from Grevesse \& Sauval (1998), there are no notable differences in the spallation elements in question by adopting Anders \& Grevesse (1989).

One important caveat is that {\sc xillver} includes only He- and H-like ions of the Fe-peak elements.  Measuring or calculating the atomic data for the neutral stages of these elements is difficult and not yet incorporated in {\sc xillver}.  The models presented here are currently the best approximation of the ionisation scenario. 

In Fig.~\ref{fig:jaspreet} (right panel), the effects of spallation on ionised ($\xi = 100\erg\cmps$) material is shown.  Several emission lines from He- and H-like species of Ti, V, Cr, and Mn become significant. Neutral stages of these elements are likely important, but not included in the current model. Ionisation generates a multitude of emission lines and modifies the energies that specific species are observed at.

\begin{figure}
\rotatebox{0}
{\scalebox{0.32}{\includegraphics{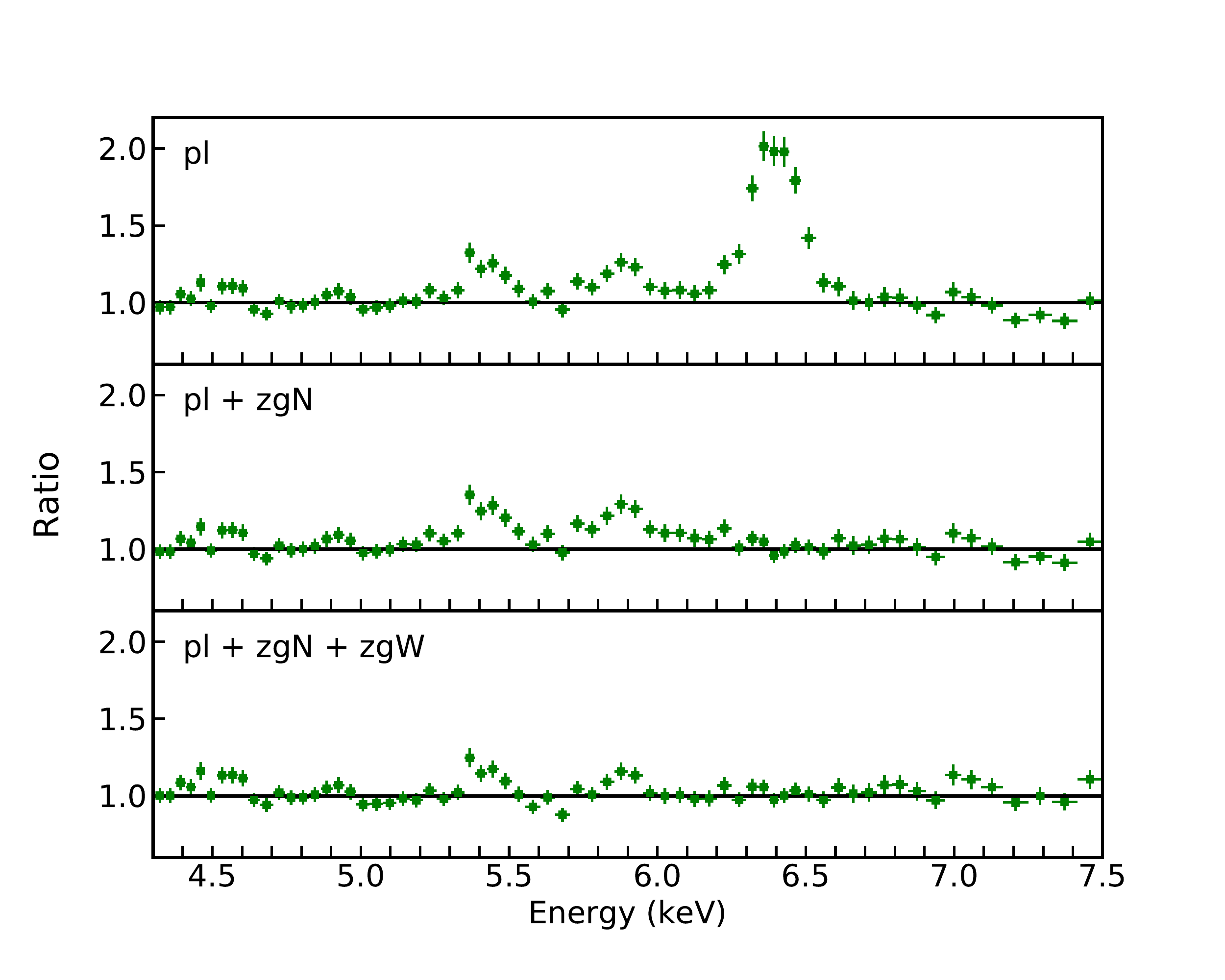}}}
\caption{
An \xmm\ (pn) simulation of a power law continuum ($\Gamma = 1.9$) being reflected from a medium with enhanced abundances of sub-iron elements from spallation as calculated in this work.  The reflection spectrum is not broadened (only narrow lines) and the reflection fraction is unity.  The $2-10\keV$ flux is $\sim 10^{-11}\ergpscmps$ and the simulation is for $100\ks$.  Top panel:  The remaining residuals in the $4.3-7.5 \keV$ band after fitting the $2.5-5$ and $7.5-10\keV$ bands with a power law (as might be done with real data).  Middle panel:  The remaining residuals after adding a narrow $6.4\keV$ Gaussian profile to the power law model.  Lower panel: The residuals that remain in the middle panel can be fitted with a broad Gaussian profile centred at $\sim5.8\keV$.  There are residuals remaining where Cr~K$\alpha$ ($5.4\keV$) and Mn~K$\alpha$ ($5.9\keV$) emission would occur that could be overlooked when modelling.
}
\label{fig:xmm}
\end{figure}

In Fig.~\ref{fig:xmm} a simulation of a typical type-I AGN spectrum as would be observed with \xmm\ is shown.  The model consists of a power law continuum and reflection with abundances modified for spallation according to Sect.~\ref{sect:spallation}.  The parameters of both components are as described above and no broadening of the reflection spectrum is considered.  This would be consistent with the reflection originating in distant material like the torus.  The power law and reflection spectrum have the same luminosity over the $0.1-100\keV$ band (i.e. the reflection fraction is unity) and the $2-10\keV$ flux is $\sim 10^{-11}\ergpscmps$.   The simulation is for a $100\ks$ exposure with the EPIC-pn.

Skibo (1997) suggested the red wing of the relativistically broadened \feka\ emission line in AGN could be attributed to the enhancement of the sub-iron spallation elements observed with CCD resolution.  As seen in the second panel of Fig.~\ref{fig:xmm}, fitting a power law and narrow Gaussian profile at $6.4\keV$ describes the simulated spectrum well, but leaves residuals between $5-6\keV$ where the enhanced Cr~K$\alpha$ and Mn~K$\alpha$ emission appear.  The addition of a single broad Gaussian profile improves these residuals (Fig.~\ref{fig:xmm}, lower panel).  If considered a priori, the spallation features are detectable with current CCD instruments.  However, a single broad profile may be considered a simpler model in such cases.

\begin{figure}
\rotatebox{0}
{\scalebox{0.32}{\includegraphics{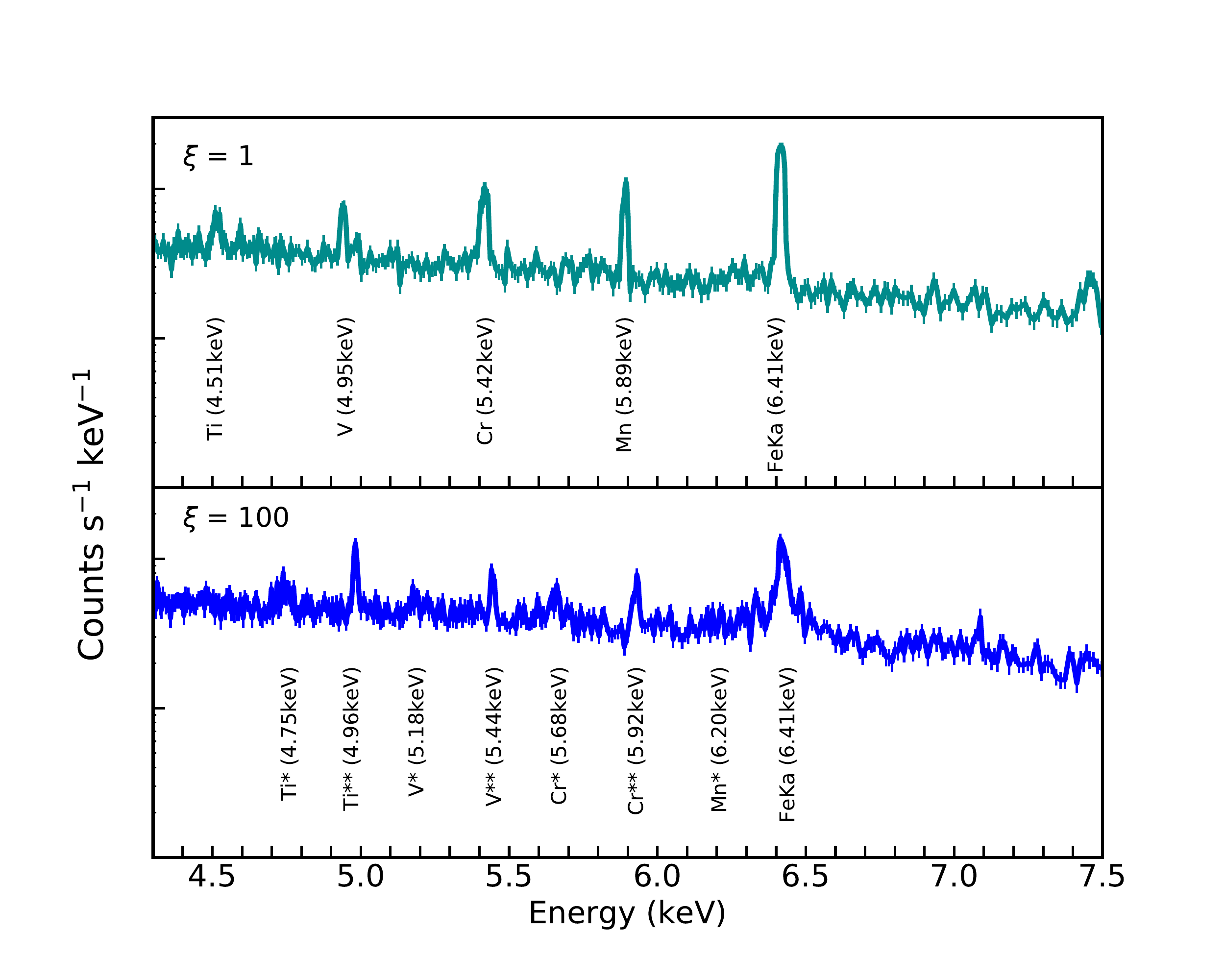}}}
\caption{
Simulation of a power law continuum ($\Gamma = 1.9$) being reflected from a medium with enhanced abundances of sub-iron elements from spallation (as in Fig.~\ref{fig:xmm}) observed with a calorimeter like the SXS on \hitomi. The $2-10\keV$ flux is $\sim 10^{-11}\ergpscmps$ and the simulation is for $250\ks$. The simulations use the average abundances calculated in this work (Section~\ref{sect:spallation}).  The case for neutral ($\xi = 1\erg\cmps$) and ionised ($\xi = 100\erg\cmps$) reflection is shown in the top and lower panel, respectively. 
}
\label{fig:hitomi}
\end{figure}

The same simulation is carried out for a $250\ks$ \hitomi\ SXS observation to examine the appearance of spallation with high spectral resolution (Fig.~\ref{fig:hitomi}, top panel).  All four of the spallation features between $\sim4.5-6\keV$ are detectable.  The calorimeter resolution could also discern ionised spallation features (Fig.~\ref{fig:hitomi}, lower panel).

\begin{figure}
\rotatebox{0}
{\scalebox{0.32}{\includegraphics{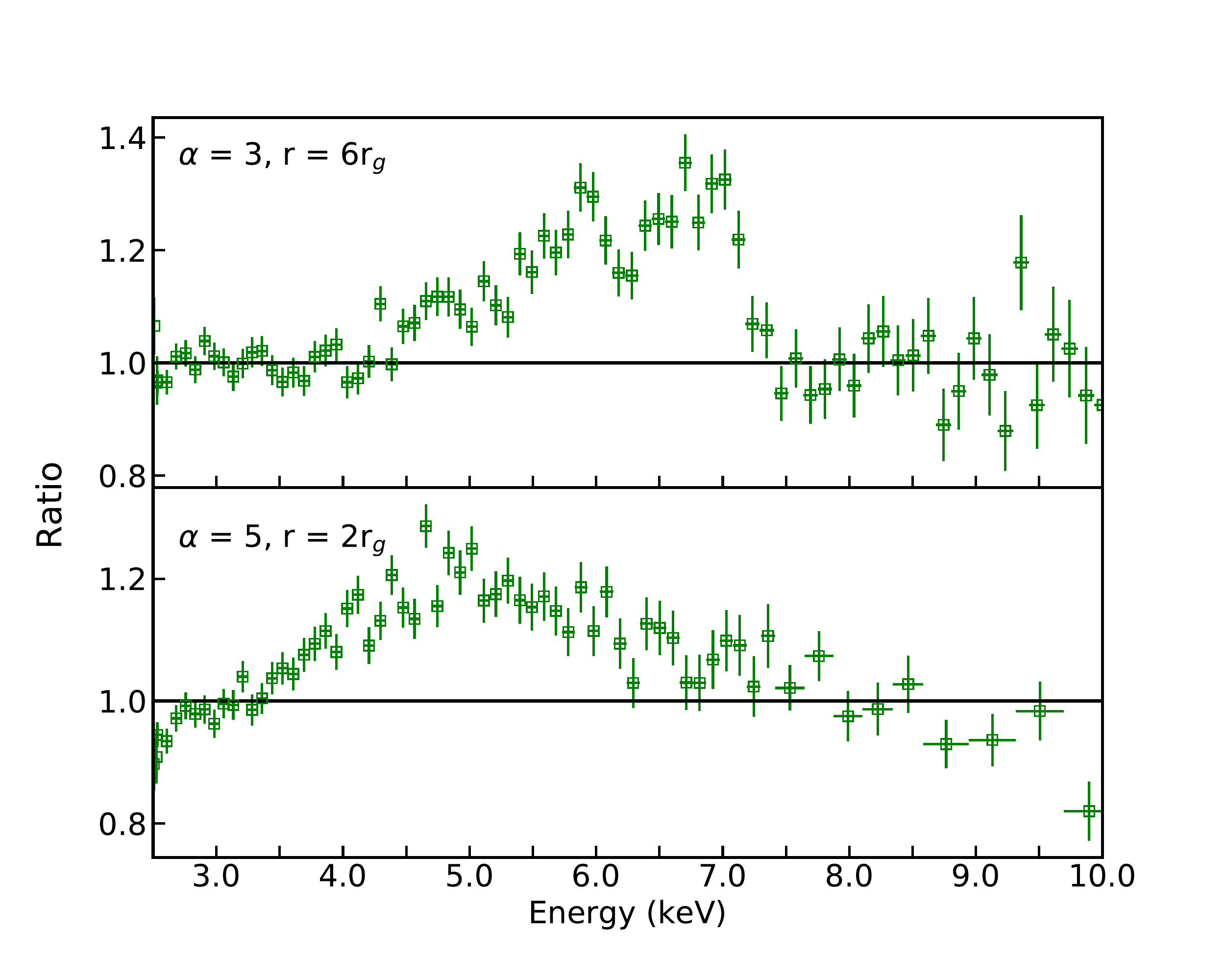}}}
\caption{
EPIC-pn simulation of a neutral reflection spectrum with enhanced sub-Fe spallation elements, blurred for motions in the accretion disc around a black hole.  In the top panel, the effects originating around a Schwarzschild black hole with the inner edge of the accretion disc fixed at $r_{in}=6\rg$ and the emissivity profile set to $\alpha=3$.  In the lower panel, the effects around a rapidly rotating black hole are simulated ($r_{in}=2\rg$ and $\alpha=5$).  
}
\label{fig:blur}
\end{figure}

The final simulation considers the effects of blurring on the spalled reflection spectrum.  Specifically, neutral reflection from material in a standard accretion disc around a Schwarzschild (Fig.~\ref{fig:blur}, top panel) and rapidly rotating black hole (Fig.~\ref{fig:blur}, lower panel) are examined.  Neutral reflection emitted from discs around zero-spin black holes provides the best-case scenario to identify sub-Fe spallation elements in the inner accretion disc.  More intense blurring or increased ionisation will substantially complicate the potential to identify spallation in the inner disc.

\section{Discussion and Conclusions} 

Cosmic ray spallation plays a role in the formation of lighter elements in the early Universe and in the Galaxy.  Whether spallation is important in AGN is debatable.  Reynolds \et (2012) argue that an unreasonably large fraction of the accretion energy would be required to generate cosmic rays of sufficient energy to produce spallation in the accretion disc.   Specifically, the resulting $\gamma$-ray emission from relativistic particles predicted by S97 for radio-quiet (jet-less) Seyferts, which are often the sources of spallation claims (e.g. Turner \& Miller 2010; Della Ceca \et 2005; Porquet \et 2004; S97),  would violate the Teng \et (2011) detection limits from {\it Fermi}. 
If there were a compact jet base in radio-quiet AGN, as has been suggested for some objects (e.g. Gallo 2018), then one might expect some spallation over a small area of the inner accretion disc.  In this case, these features would be weak and significantly blurred and likely undetectable (Fig.~\ref{fig:blur}).

It is also possible that the spallation features, along with the \feka\ line, could appear stronger in a Seyfert spectrum than predicted here.  The inner accretion disc in most Seyfert~1 galaxies is characterised with iron overabundances (e.g. Reynolds \et 2012; Garc\'ia \et 2018).  In such a scenario, the absolute strength of potential spallation features could be further enhanced, even though the relative strength compared to iron remains the same. 

Turner \& Miller (2010) suggest it is not the accretion disc that is the site of spallation, but a disc wind at some distance from the central region.  Such a ``thin'' target would require protons that are of lower energy ($E\ls200$~MeV) and only moderately relativistic ($\beta\ls0.6$), which should be possible in lower luminosity systems.  A thin target like a wind, would generate different abundances than those calculated here (Section~\ref{sect:spallation}) or by S97.  Such wind features would by subject to outflow velocities that should be detectable with future calorimeter missions (e.g. Tashiro \et 2018; Barret \et 2018).

The initial assertion by S97 that the relativistic broad line in \mcg\ and other AGN could be described by spallation of iron seen in modest-resolution \asca\ spectra is likely not valid.  Spallation features from neutral material that has not been blurred for dynamical effects would likely be distinguished with careful analysis of current CCD spectra from instruments on \xmm\ (Fig.~\ref{fig:xmm}).   Ionisation is likely to complicate detection of spallation in CCD resolution, but such features will be distinguishable in high-resolution calorimeter spectra.  Ionisation can generate a multitude of weaker lines in the spallation spectrum (Fig.~\ref{fig:jaspreet}).  Current ionisation models do not incorporate the neutral stages of Fe-peak elements, which limits the investigation of spallation in ionised material.  At the same time, the possibility of detecting features from ionised spallation elements with upcoming calorimeter missions can serve to refine needed atomic data.

The detection of the spallation of iron with calorimeter missions may also provide the opportunity to test spallation, cross-section, and abundance models themselves.  The abundances determined here are based on the S97 calculations, but with improved cross sections.  For a given line, this work and S97 predict fluxes that are comparable (e.g. within a factor of $\ls2$).  Such differences may be difficult to discern in typical X-ray data.  However, line ratios differ by more significant amounts between the two models.  For example, the for neutral reflection the line ratio $\frac{\rm Cr}{\rm V}$ is $\sim3$ as calculated in this work, but $\sim8$ according to S97.  These differences should be important to identify different spallation models, keeping in mind the effects of uncertainties (Fig.~\ref{fig:errors}), and even to distinguish between thin and thick targets (e.g. Turner \& Miller 2010).


\section*{Acknowledgments}

Many thanks to the referee for providing a helpful and stimulating report. 
We thank Dr. Lance Miller for providing the routines for proton-nucleus cross sections written by A.F. Barghouty, and
Dr. Shruti Tripathi for discussion.
MHH acknowledges the receipt of a Vanier Canada Graduate Scholarship.
J.A.G acknowledges support from NASA grant 80NSSC177K0515 and from the Alexander von Humboldt Foundation.
CSR thanks the UK Science and Technology Facilities Council for support under Consolidated Grant ST/R000867/1.




\bsp
\label{lastpage}
\end{document}